\documentclass[singlecolumn,showpacs,preprintnumbers,amsmath,amssymb]{revtex4}
\usepackage{graphicx}
\usepackage{dcolumn}
\usepackage{bm}

\begin{document}

\title{A modified proximity approach in the fusion of heavy-ions}

\author{Ishwar Dutt }
\email{idsharma.pu@gmail.com}
\author{Rajni Bansal}%
 \affiliation{Department
of Physics, Panjab University, Chandigarh -160 014, India.}


\date{\today}

\begin{abstract}
By using a suitable set of the surface  energy coefficient,
nuclear radius, and universal function, the original proximity
potential 1977 is modified. The overestimate of the data by 4\%
reported in the literature is significantly reduced. Our modified
proximity potential reproduces the experimental data nicely
compared to its older versions.
\end{abstract}


\maketitle

\section{\label{intro}Introduction}
 Recently, great theoretical and experimental
efforts are taken to studying the fusion of heavy nuclei leading
to several new phenomena including the understanding of the
formation of neutron -rich and super heavy
elements~\cite{id1,id2}. The precise knowledge of the interaction
potential between two nuclei is a difficult task and continuing
efforts are needed in this direction. This problem has been of
very active research over the last three decades and remains one
of the most widely studied subject in low-energy heavy-ion
physics~\cite{id1,id2,rkp1,blocki77,blocki81,wr94,ms2000,siwek04,wang06,
deni06}.

\par
The total interaction potential is sum of the long range Coulomb
repulsive force and short range nuclear attractive force. The
Coulomb part of the interaction potential is well-known, whereas
nuclear part is not clearly understood. A large number of efforts
have been made to giving simple and accurate forms of the nuclear
interaction
potentials~\cite{id1,id2,rkp1,blocki77,blocki81,wr94,ms2000,siwek04,wang06,deni06}.
Among such efforts, proximity potential is well known for its
simplicity and numerous applications.  Based upon the proximity
force theorem~\cite{blocki77,blocki81}, a simple formula for
ion-ion interaction potential as a function of the separation
between the surfaces of two approaching nuclei was
presented~\cite{blocki77,blocki81}.
\par
As pointed out by many authors~\cite{ms2000}, original form of the
proximity potential 1977 overestimates the experimental data by
4\% for fusion barrier heights. In a recent study involving the
comparison of 16 proximity potentials, one of us and collaborators
pointed out that proximity
 potential 1977 overestimates the experimental data by 6.7\% for
 symmetric colliding nuclei~\cite{id1}. Similar results were
 obtained for asymmetric colliding nuclei~\cite{id1}.
\par
With the passage of time, several improvement/ modifications were
made over the original proximity potential 1977 to remove the gray
part of the potential. It includes either the better form of the
surface energy coefficient~\cite{wr94} or the universal function
and/or nuclear radius~\cite{ms2000}. A careful look reveals that
these modifications/improvements are not able to explain the
experimental data~\cite{id1,siwek04}. A deep survey  also pointed
out that these technical parameters (i.e. surface energy
coefficient, nuclear radius, and universal function) were chosen
quite arbitrarily in the literature. Among them, the surface
energy coefficient is available in a large variety of forms from
time to time~\cite{id1,id2}. It affects the fusion barrier heights
and cross sections significantly~\cite{id1,id2}. Also, nuclear
radius is available in large variety of forms~\cite{id1,id2}.
These forms varies either in terms of its coefficients or either
different mass or isospin dependence. The third technical
parameter i.e, the universal function, is also parametrized in
different forms~\cite{id1,blocki77,ms2000,blocki81}.
Unfortunately, no systematic study is available in the literature,
where one can explore the role of these technical parameters in
fusion barrier positions, heights, and cross sections.
Alternatively, a best set of the above-mentioned parameters is
still missing.
\par
In the present study, our aim is to modify the original proximity
potential 1977 by using a suitable set of the above-stated
 technical parameters available in the literature. In addition, to compare the final
outcome with the huge amount of experimental data available since
last three decades. The choice of the potential and its form to be
adopted is one of the most challenging task when one wants to
compare the experimental data with theory. The present systematic
study includes the reactions with combine mass between A = 19 and
A = 294 units. In total, 390 experimentally studied reactions with
symmetric as well as asymmetric colliding partners are taken into
consideration.
Section \ref{model} describes the Model in brief,
Section \ref{result} depicts the Results and Summary is presented
in Section \ref{summary}.
\par
\section{\label{model} The Model}
The total ion-ion interaction potential  $V_{T}(r)$ between two
colliding nuclei with charges $Z_{1}$ and $Z_{2}$, center
separation $r$, and density distribution assumed spherical, and
frozen, is approximated as~\cite{ms2000}
 \begin{eqnarray}
 V_{T}(r)= V_{N}(r) + \frac{Z_{1}Z_{2}e^{2}}{r},
\label{eq:1}
\end{eqnarray}
where e is the charge unit. The above form of the Coulomb
potential is suitable when two approaching nuclei are well
separated. The nuclear part of the potential $V_{N}(r)$ is
calculated in the framework of the proximity potential 1977
\cite{blocki77} as
 \begin{equation}
V_{N}\left(r \right) = 4\pi \overline{R}\gamma
b\Phi(\frac{r-C_{1}-C_{2}}{b})  {~\rm MeV}, \label{eq:2}
\end{equation}
where $\overline{R} = \frac{C_{1}C_{2}}{C_{1}+ C_{2}}$ is the
reduced radius. Here $C_{i}$ denotes the matter radius and is
calculated using relation~\cite{ms2000}
\begin{equation}
C_{i}= c_{i}+ \frac{N_{i}}{A_{i}}t_{i}    ~~~~(i=1,2),
\label{eq:4}
\end{equation}
where $c_{i}$ denotes the half-density radii of the charge
distribution and $t_{i}$ is the neutron skin of the nucleus. To
calculate $c_{i}$, we used the relation given in
Ref.~\cite{ms2000} as
\begin{equation}
c_{i}= R_{00i}\left(1-
\frac{7}{2}\frac{b^{2}}{R_{00i}^{2}}-\frac{49}{8}\frac{b^{4}}{R_{00i}^{4}}+\cdots
\cdots   \right) ~~~~~~~(i=1,2). \label{eq:5}
\end{equation}
 Here, $R_{00}$ is the nuclear charge radius read as
\begin{equation}
 R_{00i}= 1.2332A^{1/3}_{i}
\left\{1+\frac{2.348443}{A_{i}}-0.151541\left(\frac{N_{i}-Z_{i}}{A_{i}}\right)\right\}
{~\rm fm}, \label{eq:6}
\end{equation}
where $N_{i}$ and $Z_{i}$ refer to neutron and proton contents of
target/projectile nuclei. This form of radius is taken from the
recent work of Royer and Rousseau~\cite{gr09} and is obtained by
analyzing as many as 2027 masses with N, Z $\geq$ 8 and a mass
uncertainty $\leq$ 150 keV.
 The neutron skin $t_{i}$ used in Eq. (\ref{eq:4}) is calculated according to Ref.~\cite{ms2000}.
\par
 The surface energy coefficient $\gamma$ was taken from the
work of Myers and \'Swi\c{a}tecki~\cite{ms66} and has the form
\begin{equation}
\gamma = \gamma_{0}\left[1-k_{s}\left(\frac{N-Z}{A}\right)^{2}
\right],
 \label{eq:8}
\end{equation}
where N and Z refer to the total neutrons and protons content. It
is clear from Eqs. (\ref{eq:6}) and (\ref{eq:8}) that both nuclear
radius as well as surface energy coefficient depend on the
relative neutron excess. In the above formula, $ \gamma_{0}$ is
the surface energy constant and $k_{s}$ is the surface-asymmetry
constant. Both constants were first parameterized by Myers and
\'Swi\c{a}tecki~\cite{ms66} by fitting the experimental binding
energies. The first set of these constants yielded values
$\gamma_{0}$ and $k_{s}=1.01734 ~\rm
 ~MeV/fm^{2}$ and 1.79, respectively. In original proximity version, $\gamma_{0}$ and $k_{s}$
 were taken to be $0.9517 ~\rm
 ~MeV/fm^{2}$ and 1.7826~\cite{ms67}, respectively. Later on, these
values were  revised in a large variety of forms depending upon
the advancement in the theory as well in
experiments~\cite{id1,id2}. In total, 14 such coefficients are
highlighted in Ref.~\cite{id2} and the role of extreme 4 sets is
analyzed deeply. Out of them, two best sets of surface energy
coefficients are stressed. In the present study, we shall restrict
to the latest set of $\gamma$ values i.e. $\gamma_{0}$ =$1.25284
{~\rm MeV/fm^{2}}$ and $k_{s}$ = $2.345$ presented in
Ref~\cite{id2}. This particular set of values were obtained
directly from a least-squares adjustment to the ground-state
masses of 1654 nuclei ranging from $^{16}$O to $^{263}$106 and
fission-barrier heights~\cite{mn95}.

\par
  The universal function $\Phi(\frac{r-C_{1}-C_{2}}{b})$ used in Eq.
(\ref{eq:1}) has been derived by several authors in different
forms~\cite{blocki77,ms2000,blocki81}. In original proximity
potential, $\Phi(\frac{r-C_{1}-C_{2}}{b})$ was parametrized in the
cubic-exponential form~\cite{blocki77}
\begin{equation}
\Phi \left(\xi \right)= \left\{
\begin{array}{l}
-\frac{1}{2} \left(\xi- 2.54 \right)^{2}-0.0852\left(\xi- 2.54
\right)^{3},~ \mbox{ for $\xi \leq 1.2511 $ },    \\
-3.437\exp \left(-\frac{\xi}{0.75} \right),~~~~~~~~~~~~~~~~~~~~~~~
\mbox{ for $\xi \geq  1.2511 $ },
\end{array}
\right. \label{eq:9}
\end{equation}
with $\xi$ = $(r - C_{1} - C_{2}$)/$b$. The surface width $b$
(i.e.~$b=\frac{\pi}{\sqrt{3}}a ~{\rm with}~ a=0.55 ~\rm fm)$ has
been evaluated close to unity. We labeled this universal function
as $\Phi$-${1977}$.
\par
Later on,  Blocki et al.,~\cite{blocki81} modified the above form
as
\begin{equation} \Phi \left(\xi \right)= \left\{
\begin{array}{l}
-1.7817+ 0.9270\xi +0.143\xi^{2}-0.09\xi^{3},~~~~~~ \mbox{ for $\xi \leq 0.0 $ },    \\
-1.7817+ 0.9270\xi +0.01696\xi^{2}-0.05148\xi^{3},\\
~~~~~~~~~~~~~~~~~~~~~~~~~~~~~~~~~~~~~~~~~~~ \mbox{ for $0.0 \leq \xi \leq 1.9475 $ },    \\
-4.41\exp \left(-\frac{\xi}{0.7176} \right),~~~~~~~~~~~~~~~ \mbox{
for $\xi \geq  1.9475 $ }.
\end{array}
\right. \label{eq:10}
\end{equation}
In the present study, we use this form of universal function and
marked it as $\Phi$-${1981}$. By using the above stated
parameters, we construct a new proximity potential and labeled as
Prox 2010. Along with the above modified form, we shall also use
the original proximity potential 1977~\cite{blocki77} and its
recently modified form proximity potential 2000~\cite{ms2000}. We
labeled them as Prox 1977 and Prox 2000, respectively.


%
\section{\label{result}Results and Discussions}
By using the above new version of the proximity potential (Prox
2010) along with its older versions (i.e. Prox 1977 and Prox
2000), fusion barriers are calculated for 390 reactions by using
the conditions:
 \begin{equation}
\frac{dV_T(r)}{dr}|_{r=R_{B}} = 0,~~ {\rm{and}} ~~
\frac{d^{2}V_T(r)}{dr^{2}}|_{r=R_{B}} \leq 0. \label{eq:11}
\end{equation}
The height of the barrier and position is marked, respectively, as
$V_{B}$ and $R_{B}$.

\par
As one see from the preceding section, three factors govern the
success of proximity potential are (i) the surface energy
coefficient, (ii) the universal function, and (iii)  nuclear
radius. We analyzed the literature very carefully and found that
the latest information on these three factors can shape the new
proximity potential. Recently, the role of surface energy
coefficient stated above is studied in detail in Ref~\cite{id2}.
 As for as radius is concern, we shall
restrict to its latest form given in Ref.~\cite{gr09}. However,
the role of the third parameter i.e., the universal function in
fusion barriers is analyzed in Fig. 1. Here, we display $\Delta
V_{B}~(\%)$ and $\Delta R_{B}~(\%)$ defined as
\begin{equation}
\Delta V_{B}~(\%)= \frac{V_{B}^{theor} -
V_{B}^{expt}}{V_{B}^{expt}}\times 100, \label{eq:12}
\end{equation}
and
\begin{equation}
\Delta R_{B}~(\%) = \frac{R_{B}^{theor}-
R_{B}^{expt}}{R_{B}^{expt}}\times 100, \label{eq:13}
\end{equation}
as a function of $Z_{1}Z_{2}$ using two sets of above mentioned
universal functions [Eqs. (\ref{eq:9}) and (\ref{eq:10})]. It is
clear from the figure that deviations are significantly reduced by
using $\Phi$-${1981}$ compared to its original form
$\Phi$-${1977}$. The universal function $\Phi$-${1981}$ reduces
the average deviation over 390 reactions by 1 \% for fusion
barriers. The experimental values are taken directly from the
literature~\cite{id1,id2,ms2000}. Actually, it is clear from the
literature that no experiment can extract information about the
fusion barriers directly. All experiments measure the fusion
differential cross sections and then with the help of a
theoretical model, one can extract the fusion barriers.
\par
In Fig. 2, we display the theoretical fusion barrier heights
$V_{B}^{theor}$ (MeV) and positions $R_{B}^{theor}$ (fm) verses
the corresponding experimental values.  We note from the figure
that Prox 2010 potential reproduces the experimental fusion
barrier heights within 1.4\%. This result is in close agreement
with other recently parametrized potentials presented in
Ref.~\cite{id1}. However, the original form of the proximity
potential 1977 presented in Ref.~\cite{id1} overestimates the data
by 6.7\% for symmetric colliding nuclei. However, the fusion
barrier positions show some scattering from the central line
(marked by shaded area). This scattering may be due to the
variation in the experimental setups and theoretical method one
used to extract these values~\cite{Aljuwair84,Trotta01}.
\par
We quantify our outcome in Figs. 3 and 4. In Fig. 3, the
percentage deviations between the theoretical  and experimental
values are presented. The original proximity potential 1977 (Prox
1977) along with its recently modified form (Prox 2000) are also
displayed. We note from the upper panel of Fig. 3 that Prox 2010
potential on average gives better results compared to its older
versions for fusion barrier heights. However, slight  deviations
are visible for fusion barrier positions. This may be due to the
fact that in the proximity potential Prox 2010, we use the value
of surface energy coefficient that gives stronger attraction
compared to one used in Prox 1977 and Prox 2000 potentials.
Therefore, in Prox 2010 potential, the counterbalance between the
repulsive Coulomb and attractive nuclear part of the interaction
potential occurs at larger distances, and hence it pushes the
barrier outwards. The fusion barrier heights are reproduced within
$\pm~5\%$ on average. On the other hand, fusion barrier positions
reproduced the experimental values within $\pm~10\%$. Especially
for the heavier colliding nuclei, we see that Prox 2010 potential
reproduces the data much better on the average compared to other
versions. For lighter nuclei, however, small scattering is
visible. This could also be due to the uncertainty in the radius
of the lighter colliding nuclei. In Figs. 1-4, only 155 reactions
are displayed to maintain the clarity. The average deviation for
the fusion barrier heights over 390 reactions is 0.77 \% using our
modified potential Prox 2010, whereas Prox 1977, and Prox 2000
give 3.99 \%, and 4.45 \%, respectively. This shows that our
modified proximity explains the experimental data nicely.
\par
In Fig. 4, we display the difference  between the theoretical and
experimentally extracted fusion barriers. We further note that
Prox 2010 potential gives better results. The difference
especially for the heavy systems is significantly reduced. This
was the problem with original as well as its recently modified
form as pointed out by several authors~\cite{ms2000,siwek04}. It
is clear from Figs. 3 and 4, that Prox 2010 potential is able to
reproduce the experimental data much better than its older
versions.  The small difference is not significant because of the
uncertainties in the analysis of the experimental data.
\par
Finally, we test our newly modified proximity potential Prox 2010
on fusion probabilities. In Fig. 5, we display the fusion cross
sections $\sigma_{fus}$ (in mb) as a function of the
center-of-mass energy $E_{c.m.}$ (MeV) for the reactions of
$^{26}Mg+^{30}Si$~\cite{Morsad90}, $^{16}O+^{46}Ti$~\cite{Neto90},
$^{48}Ca+^{48}Ca$~\cite{Stefanini09},
$^{12}C+^{92}Zr$~\cite{newton01},
$^{40}Ca+^{58}Ni$~\cite{sikora79}, and
$^{16}O+^{144}Sm$~\cite{leigh95}. The fusion cross sections are
calculated using well known Wong  model~\cite{wg72}. The older
versions of proximity potentials that is, Prox 1977 and Prox 2000
are also displayed. It is clearly visible from the figure that
Prox 2010 potential is in good agreement, whereas, its older forms
are far from the experimental data. We further note that Prox 1977
and Prox 2000 potentials show similar results. It means that no
improvements is seen in Prox 2000 potential as was claimed in
Ref.~\cite{ms2000}.
\par
\section{\label{summary}Summary}
In the present study, we present a best set of the surface energy
coefficient, the nuclear radius, and the universal function
available in the literature. We find that these parameters which
were used quite arbitrarily in past years affect the fusion
barrier heights, positions, and cross sections significantly. By
using the above set of parameters, a new proximity potential is
constructed. Our newly constructed proximity potential Prox 2010
reproduces the fusion barriers and cross sections better than its
earlier
versions.  \\


\newpage

\begin{figure}
\centering
\includegraphics* [scale=0.40] {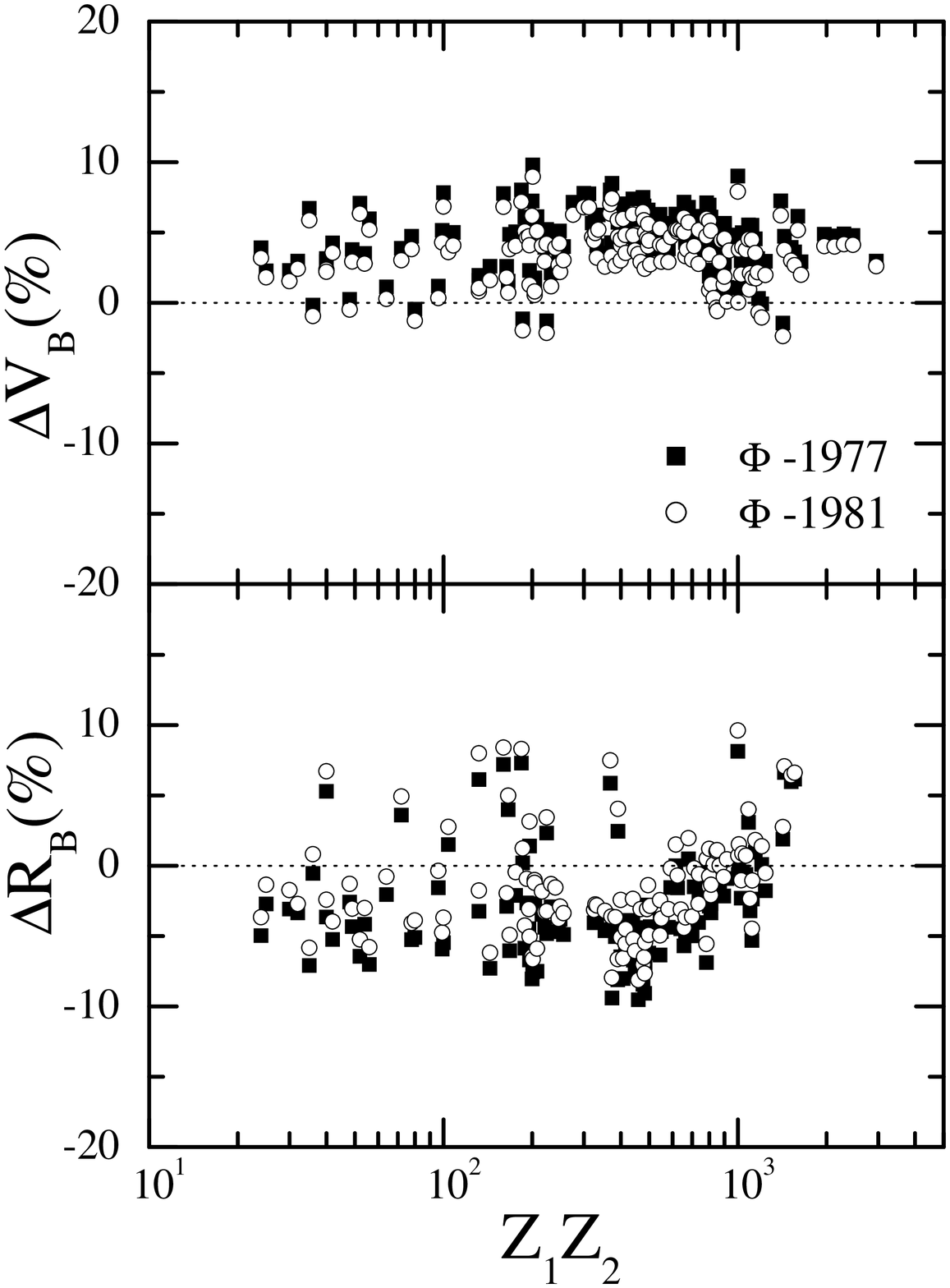}
 \caption {The percentage deviation $\Delta V_{B}~
(\%)$ and $\Delta R_{B} ~(\%$) as a function of
 $Z_{1}Z_{2}$ using two different sets of universal functions [Eqs. (\ref{eq:9}) and (\ref{eq:10})] implemented in the original proximity potential
 Prox 1977. The experimental values are taken from Refs.~\cite{id1,id2,ms2000}.}
\end{figure}
\begin{figure}
\centering
\includegraphics* [scale=0.40] {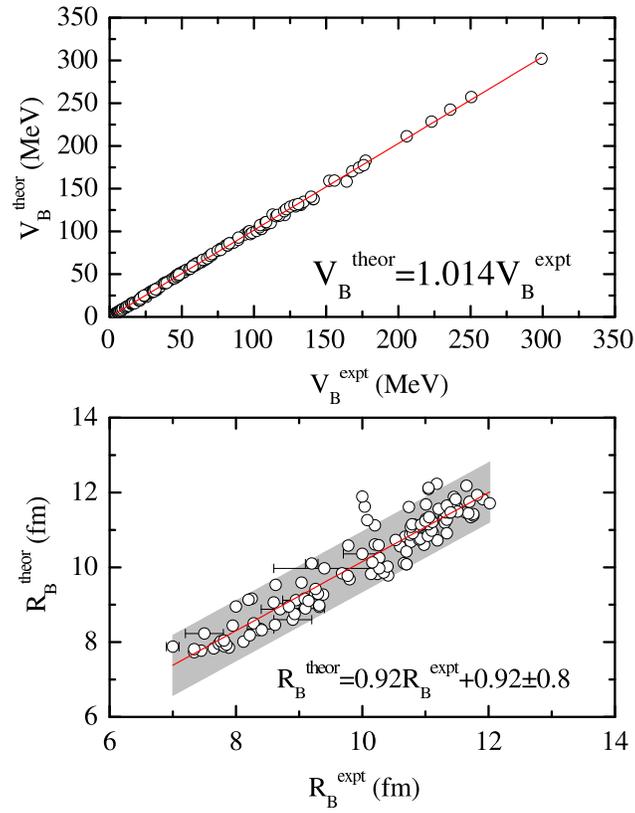}
\caption {The fusion barrier heights $V_{B}$ (MeV)
and positions barriers $R_{B}$ (fm)
  as a function of the corresponding experimental values using our modified proximity potentials Prox 2010.
  The experimental values are taken from Refs.~\cite{id1,id2,ms2000}}
\end{figure}
\begin{figure}
\centering
\includegraphics* [scale=0.40] {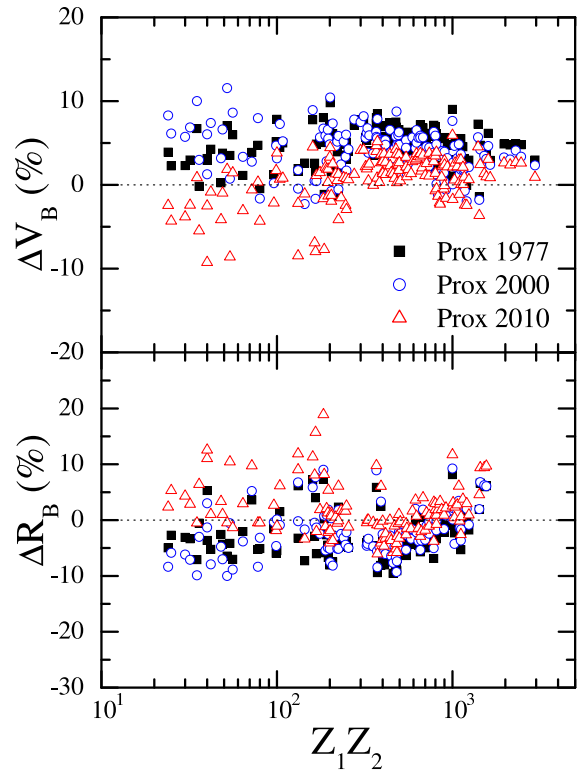}
\caption {The same as Fig.1, but for different
older proximity potentials along with our modified form i.e., Prox
1977, Prox 2000, and Prox 2010, respectively.}\label{fig3}
\end{figure}
\begin{figure}[!t]
\begin{center}
\includegraphics*[scale=0.4] {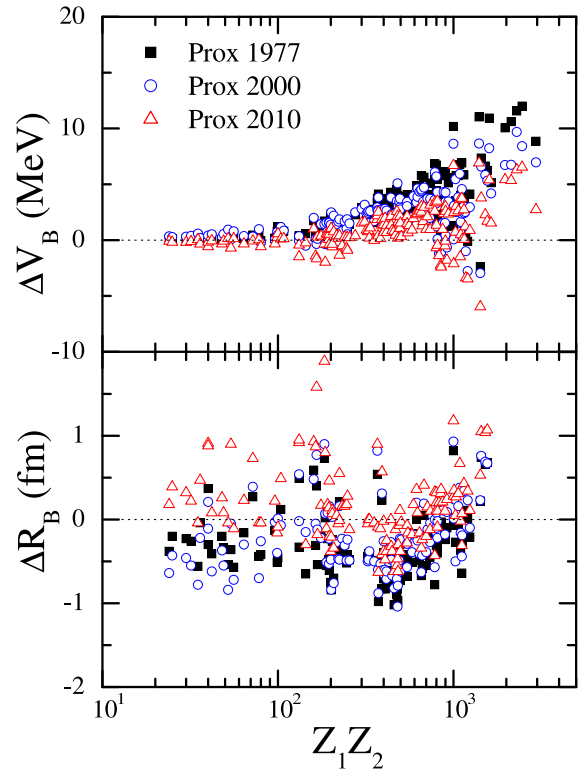}
 \caption {The variation of $\Delta V_{B}$~$(=
V_{B}^{theor}-V_{B}^{expt})$ and $\Delta R_{B}$~$(=
R_{B}^{theor}-R_{B}^{expt})$ as a function of
 $Z_{1}Z_{2}$ using Prox 1977, Prox 2000, and Prox 2010 potentials. The experimental values are taken from Refs.~\cite{id1,id2,ms2000}.}
\end{center}
\end{figure}
\begin{figure}[!t]
\centering
\includegraphics* [scale=0.42]{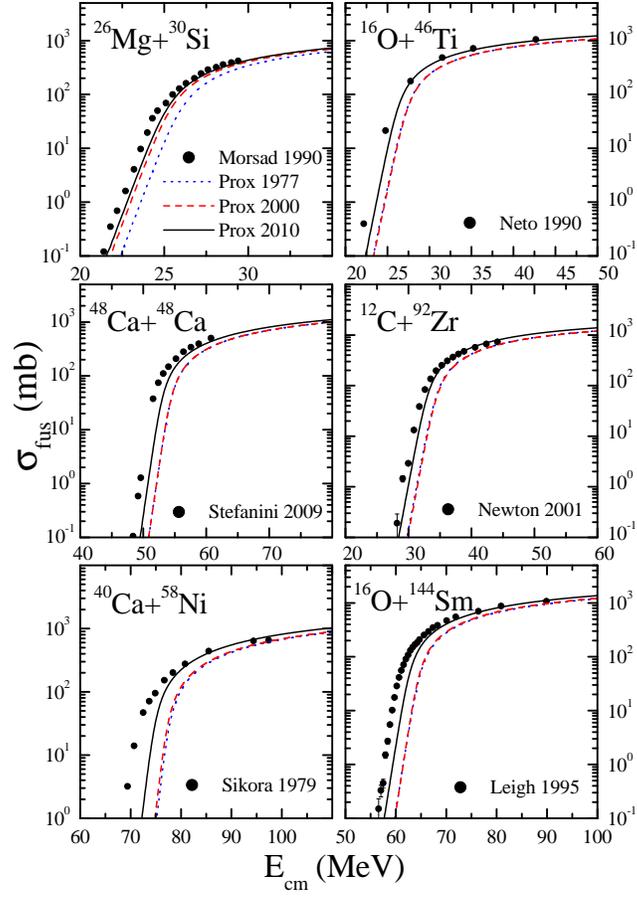}
\caption {(Color online) The fusion cross sections
$\sigma_{fus}$ (mb) as a function of center-of-mass energy
$E_{c.m.}$ using older versions of proximity potential (Prox 1977
and Prox 2000) along with new version (Prox 2010).
 The experimental data are taken from Morsad 1990~\cite{Morsad90}, Neto
1990~\cite{Neto90}, Stefanini~\cite{Stefanini09}, Newton
2001~\cite{newton01}, Sikora 1979~\cite{sikora79}, and Leigh
1995~\cite{leigh95}.}\label{fig4}
\end{figure}

\end{document}